\documentclass[12pt]{iopart}
\usepackage{ngerman}
\usepackage{graphicx}
\usepackage{mathrsfs}
\usepackage{bbm}
\usepackage[english]{babel}
\begin{document}
\title{Sticky physics of joy: On the dissolution of spherical candies }
\date{\today}
\author{Andreas Windisch}
\ead{andreas.windisch@uni-graz.at}
\address{Institut f\"ur Physik, Universit\"at Graz, Universit\"atsplatz 5, 8010 Graz, Austria}
\author{Herbert Windisch}
\ead{herbert.windisch@medunigraz.at}
\address{Retired, Institut f\"ur Biophysik, Medizinische Universit\"at Graz, Harrachgasse 21, 8010 Graz, Austria}
\author{Anita Popescu}
\ead{popescu@sbox.tugraz.at} 
\address{Institut f\"ur Statistik, Technische Universit\"at Graz, Kopernikusgasse 26, 8010 Graz, Austria}
\begin{abstract}
Assuming a constant mass-decrease per unit-surface and -time we provide a very simplistic model for the dissolution process of spherical candies. The aim is to investigate the quantitative behavior of the dissolution process throughout the act of eating the candy. In our model we do not take any microscopic mechanism of the dissolution process into account, but rather provide an estimate which is based on easy-to-follow calculations. Having obtained a description based on this calculation, we confirm the assumed behavior by providing experimental data of the dissolution process. Besides a deviation from our prediction caused by the production process of the candies below a diameter of $2\ mm$,  we find good agreement with our model-based expectations. Serious questions on the optimal strategy of enjoying a candy will be addressed, like whether it is wise to split the candy by breaking it with the teeth or not.
\end{abstract}
\submitto{Physics Education}
\maketitle
\section{\label{sec:intro}Introduction}
\begin{figure}[h]
\centering
\includegraphics[width=2.5cm]{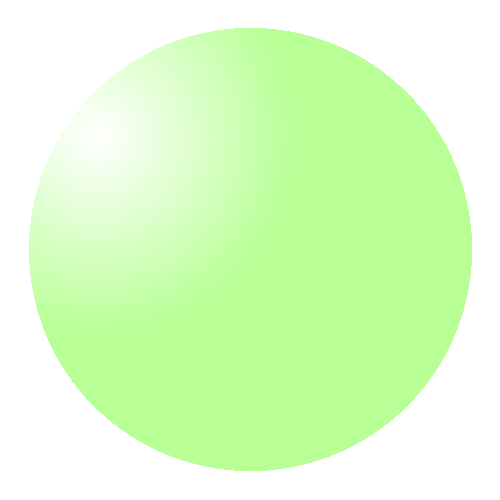}
\caption{A spherical sugar-made homogeneous and isotropic candy. How long does it take till the candy is dissolved, and \textit{how} does it vanish?} 
\label{Fig1}
\end{figure}

Let us consider a 3-dimensional spherical sugar-made candy of given mass $m_0$, radius $r_0$ and density $\rho$ as depicted in Figure \ref{Fig1}. The mass-transfer rate $c$ is assumed to be constant, i.e. within an infinitesimal time-step, the total amount of mass that is dissolved depends on the surface area of the candy at this very instant only.
If one enjoys such a candy, the dissolution process kicks in as soon as the candy is embedded in the saliva-reservoir of the oral cavity of the lucky connoisseur.  However, the time of joy due to tastiness is quite finite. In order to provide a rough estimate for the time the candy survives this for it hostile-to-live environment, let us try to \textit{quantify} this problem. Clearly, the mass decreases due to the dissolution process. The constant mass-transfer rate decreases the surface area, which in turn reduces the overall reduction of mass. Hereby we assume that the candy keeps its shape throughout the whole dissolution process. As long as the shape of the candy stays the same, we can use the same formulas for its surface, volume, mass and radius throughout the dissolution process. As revealed by the actual measurement of spherical candy dissolution, the assumption of the candy keeping its shape is fulfilled quite well, leading to good predictions with this model. 
\section{\label{sec:calc}A simplistic model}
The desired quantity is the function that describes the decrease of the mass of the candy in time,
\begin{equation}
\frac{dm}{dt}=c \cdot s(m),
\label{eq2}
\end{equation}
where $s(m)$ is the surface area of the candy for a given mass $m$ and $c\ (c\in\mathbbm{R},\ c<0)$ is the mass-transfer rate which is assumed to be constant. That is, the change of the mass in time is proportional to the surface area of the candy expressed in terms of its mass via the mass-transfer rate. In order to proceed, we have to express the surface area of the candy as a function of a given (time dependent) mass, which can be done due to the assumption of constant density. By using the following equations
\begin{eqnarray}
\label{eq3}
v(m) &=& \frac{4\pi}{3}\ r(m)^3\\
\label{eq4}
s(m) &=& 4\pi\ r(m)^2\\
\label{eq6}
\rho &=& \frac{m}{v}\\
\label{eq5}
r(m) &=& \sqrt[3]{\frac{3m}{4\pi\rho}}
\end{eqnarray}
for the volume, surface, density and radius of the candy at an instant, we can express the surface area of the 3-sphere as
\begin{equation}
s(m)=4\pi\left(\frac{3m}{4\pi\rho}\right)^\frac{2}{3}.
\label{eq7}
\end{equation}
Plugging equation (\ref{eq7}) into (\ref{eq2}) we have
\begin{equation}
\frac{dm}{dt}=-|c|\cdot 4\pi\left(\frac{3m}{4\pi\rho}\right)^\frac{2}{3}.
\label{eq8}
\end{equation}
Equation (\ref{eq8}) is a homogeneous first-order nonlinear ordinary differential equation. By solving it we obtain the mass as a function of time. All relevant parameters of the candy (volume, surface, radius) follow then directly by re-substituting the mass at a given time into the equations (\ref{eq3}), (\ref{eq4}) and (\ref{eq5}) respectively. We employ $m(t=0)=m(0)=m_0$ as initial condition, which is the mass of the yet untouched candy. The (real) solution of (\ref{eq8}) can be obtained easily by separation of variables:
\begin{eqnarray}
\label{eq8a}
\frac{dm}{dt}&=&-\underbrace{|c|\cdot 4\pi\left(\frac{3m}{4\pi\rho}\right)^\frac{2}{3}}_{=A}m^\frac{2}{3},\\
\label{eq8b}
\frac{dm}{dt}&=&-A\ m^\frac{2}{3},\\
\label{eq8c}
\int_{m_0}^m\frac{d\tilde{m}}{\tilde{m}^\frac{2}{3}}&=&-A\int_0^t d\tilde{t},\\
\label{eq8d}
3 m^{\frac{1}{3}}-3 m_0^\frac{1}{3} &=&-A\ t,\\
\label{eq8e}
m(t) &=& \left(m_0^{\frac{1}{3}}-\frac{A}{3}t\right)^3,\\
\label{eq8f}
m(t)&=&m_0-A\ m_0^{2/3}\ t+\frac{1}{3}\ A^2\ m_0^{1/3}\ t^2-\frac{1}{27}\ A^3\ t^3.
\label{eq9}
\end{eqnarray}
This solution now gives access to all relevant parameters of the candy. Hereby, the parameters (radius, surface, volume) expressed in terms of the mass inherit their time dependence from the mass. Already by looking at equation (\ref{eq8}), we can tell that the dissolution of the candy-sphere's mass does \textit{not} occur exponentially, since equation (\ref{eq8}) is not of the type $\dot m\propto m(t)$.\par
Furthermore, equation (\ref{eq8}) possesses two additional solutions which are complex. They occur due to the root on the right-hand-side of the equation and are complex conjugate to each other. They don't seem to hold any relevant information, thus we concentrate on the real solution only, where we restrict the domain to positive or vanishing mass values.\par
\begin{figure}
\centering
\includegraphics[width=12cm]{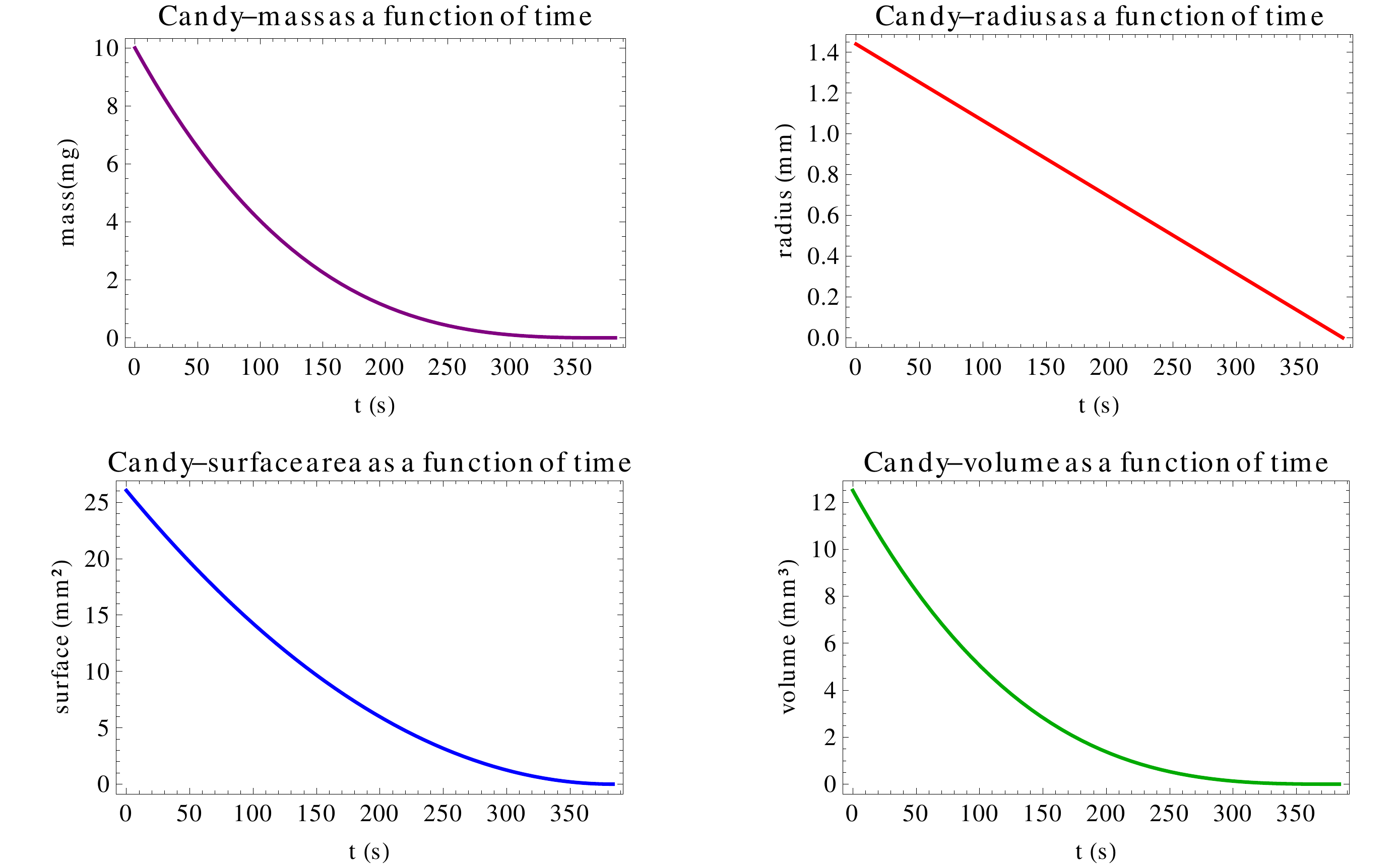}
\caption{The mass, radius, surface area and volume of a spherical candy with an initial mass $m_0=10\ mg$, density $\rho=0.8\ \frac{mg}{mm^3}$ and a mass-transfer rate $c=-0.003\ \frac{mg}{s\cdot mm^2}$.}
\label{fig2}  
\end{figure}

Figure \ref{fig2} shows the solution (\ref{eq9}) for a given set of arbitrarily chosen parameters. The Figure shows that the radius of the candy vanishes linearly in time. The solution (\ref{eq8e}) is a $3^{rd}$-order polynomial in $t$ of the generic form
\begin{equation}
m(t)=(a-k\ t)^3,
\label{eq9a}
\end{equation}
with $a=m_0^{\frac{1}{3}}$ and $k=\frac{A}{3}$. Since $r$  in equation (\ref{eq5}) is proportional to the cubic root of $m$, it follows that the radius has to be proportional to $a-k\ t$, i.e. the radius vanishes linearly in time.\par 
The radius (or to be more precise, the diameter) of the sphere is an experimentally excellent accessible quantity, thus we will use the prediction of the linear dissolution in terms of the diameter and seek experimental proof for this statement. 
\section{\label{sec:meas}Methods}
In the previous section we proposed that within our idealized model the candy vanishes linearly in time when considering the radius or diameter to describe the state of the candy in an instant. To prove this behavior we constructed an experimental setup, allowing us to determine the diameter of the candy without interfering with the dissolution process. Thus the candy dissolves under conditions close to our theoretical assumptions.\par
\begin{figure}
\centering
\includegraphics[width=8cm]{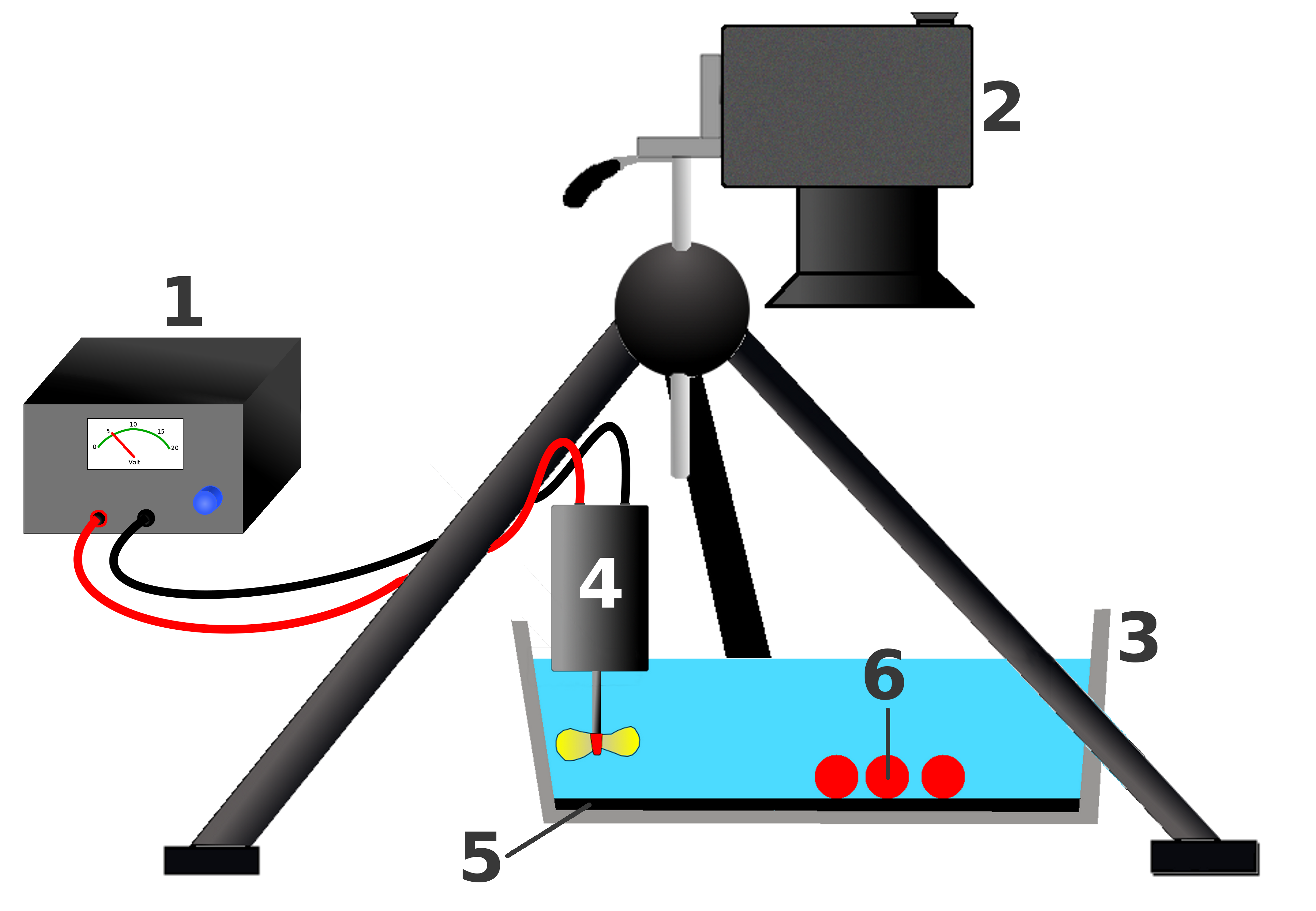}
\caption{Schematic drawing of the experimental setup: 1 power supply, 2 digital camera, 3 a bowl containing water, 4 stirring motor, 5 black rubber floor, 6 three candy samples.}
\label{fig3}  
\end{figure}
We used ordinary tap water in a bowl as environment for the dissolution. The pH of water at room-temperature is roughly 7, which corresponds to the pH of saliva \cite{pH}. Adequate samples in form of homogeneous spherical sugar-made candies are provided by \cite{candy}. The experimental setup is depicted in Figure \ref{fig3}. A power supply drives a small electric motor which stirs the water contained in the bowl slightly. Without stirring, the candies produce clouds of saturation in their immediate vicinity, such that additional diffusion effects become dominant. Directly above the water bowl a digital camera is mounted on a tripod \cite{nikon}. The camera is equipped with the capability of taking series of photos in fixed time intervals. By taking the time interval to be one minute, and by using the date and time stamp feature of the camera, we get a series of pictures showing the candies in the water bowl in a top-down view in time-distances of one minute. The bottom of the water-bowl has been covered by black foam rubber to enhance the contrast for the candies. Once the candies are dissolved, the frames taken throughout the measurement are processed on a computer. By taking a 50-Euro cent coin as a reference, a distance-per-pixel calibration value was determined. The setup even allows for unattended measurements.\par
\begin{figure}
\centering
\begin{tabular}{ c c}	
		\includegraphics[width=4.5cm]{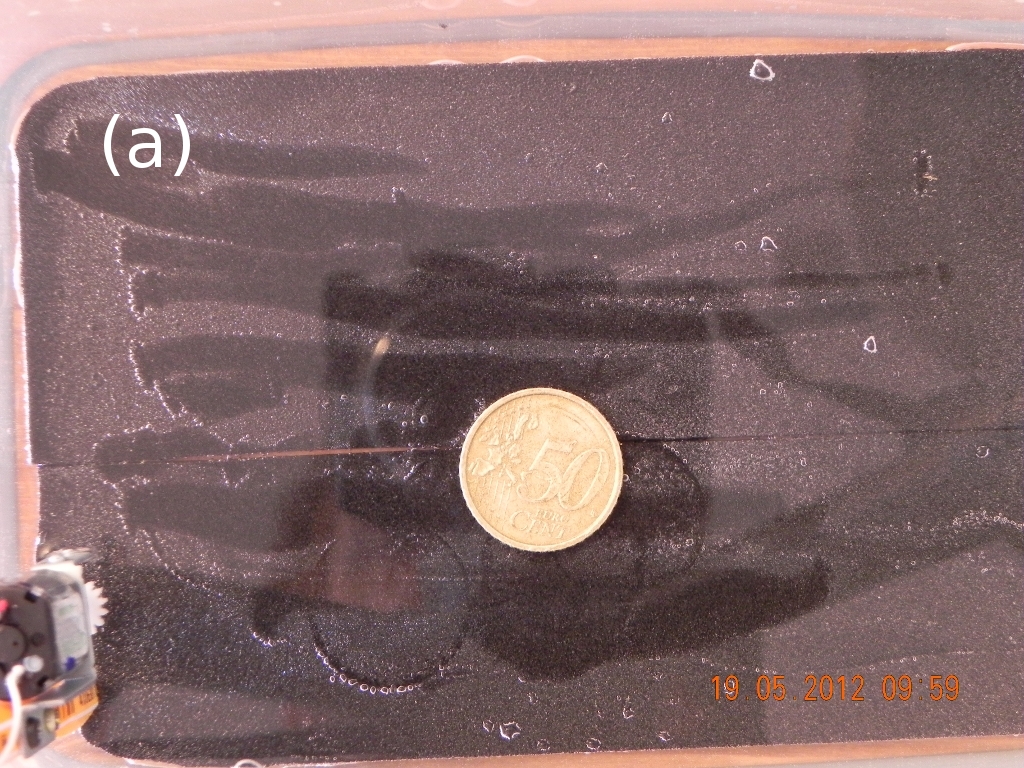}
	&	\includegraphics[width=4.5cm]{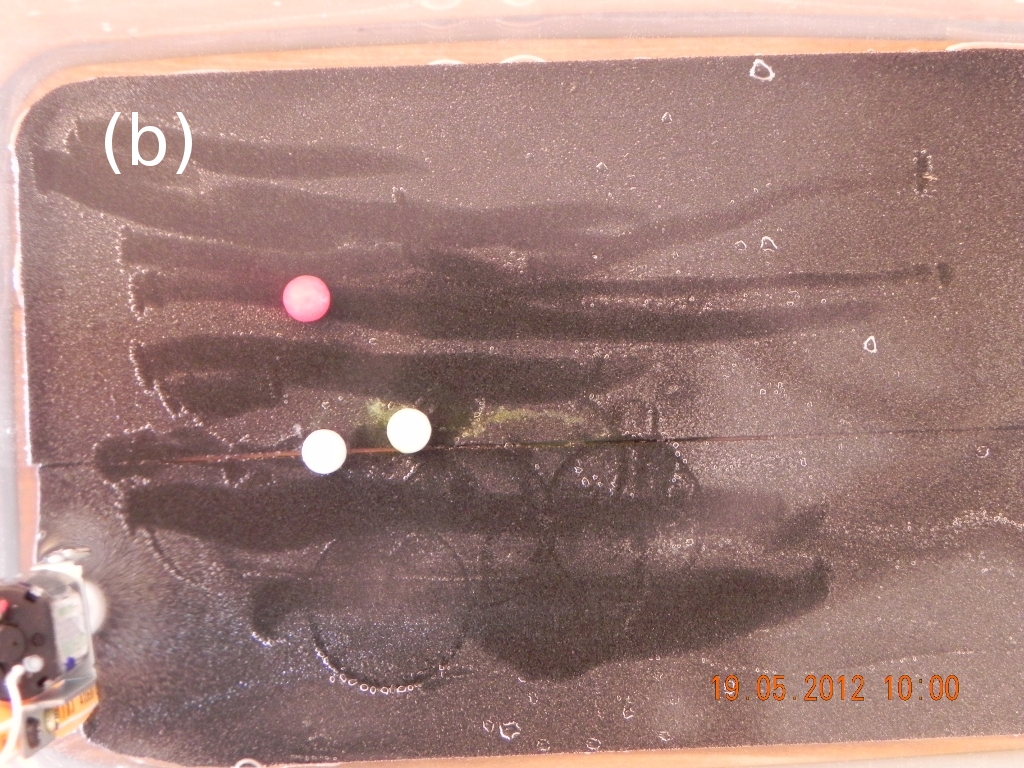}\\
		\includegraphics[width=4.5cm]{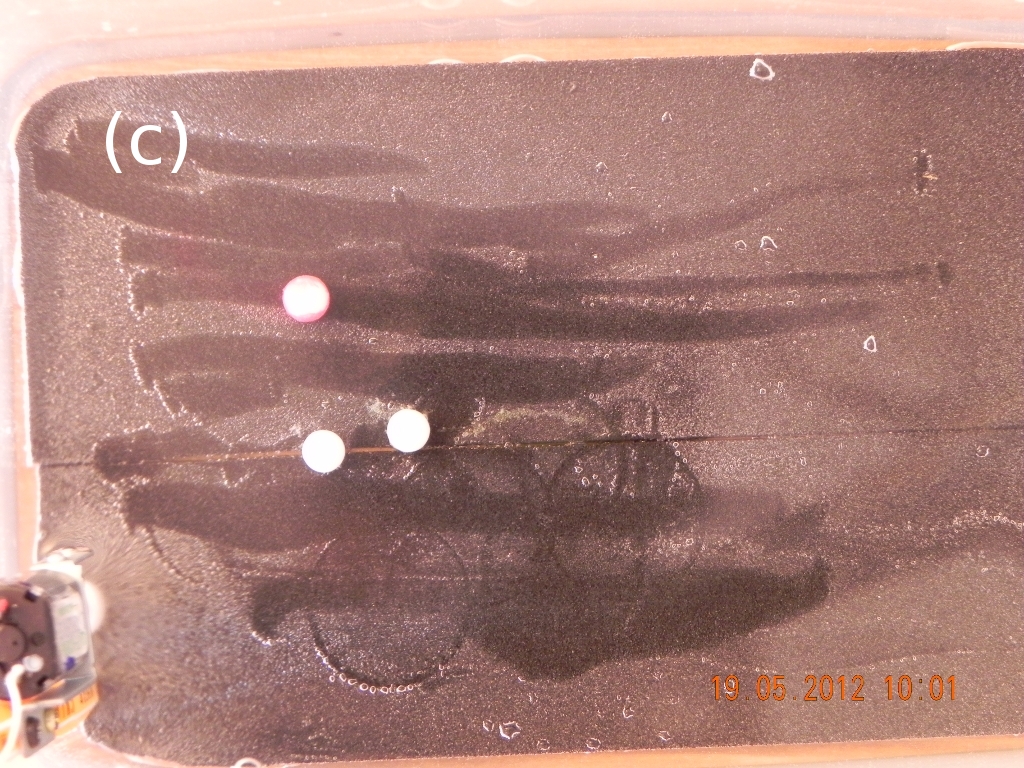}
	&	\includegraphics[width=4.5cm]{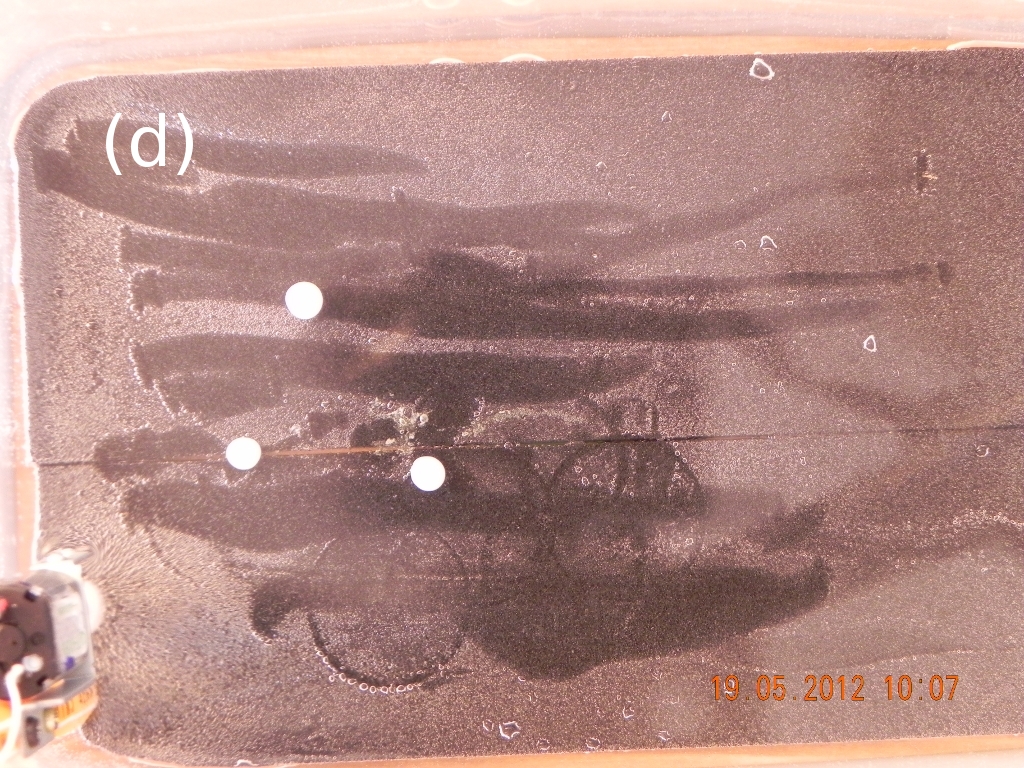}\\
		\includegraphics[width=4.5cm]{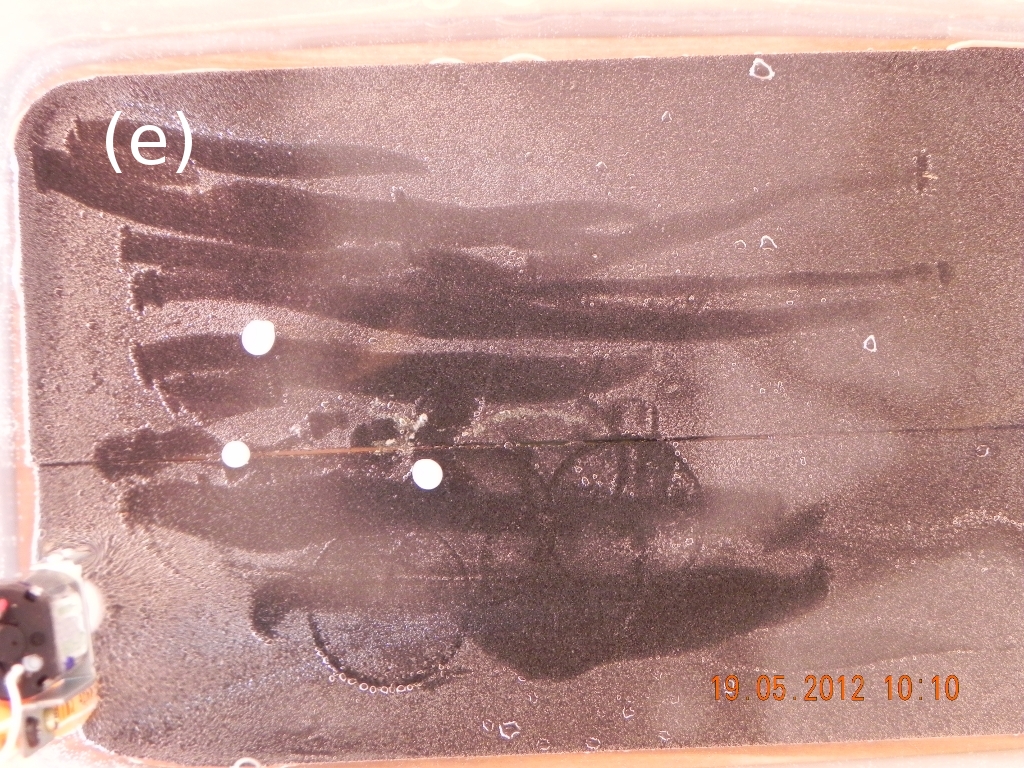}
	&	\includegraphics[width=4.5cm]{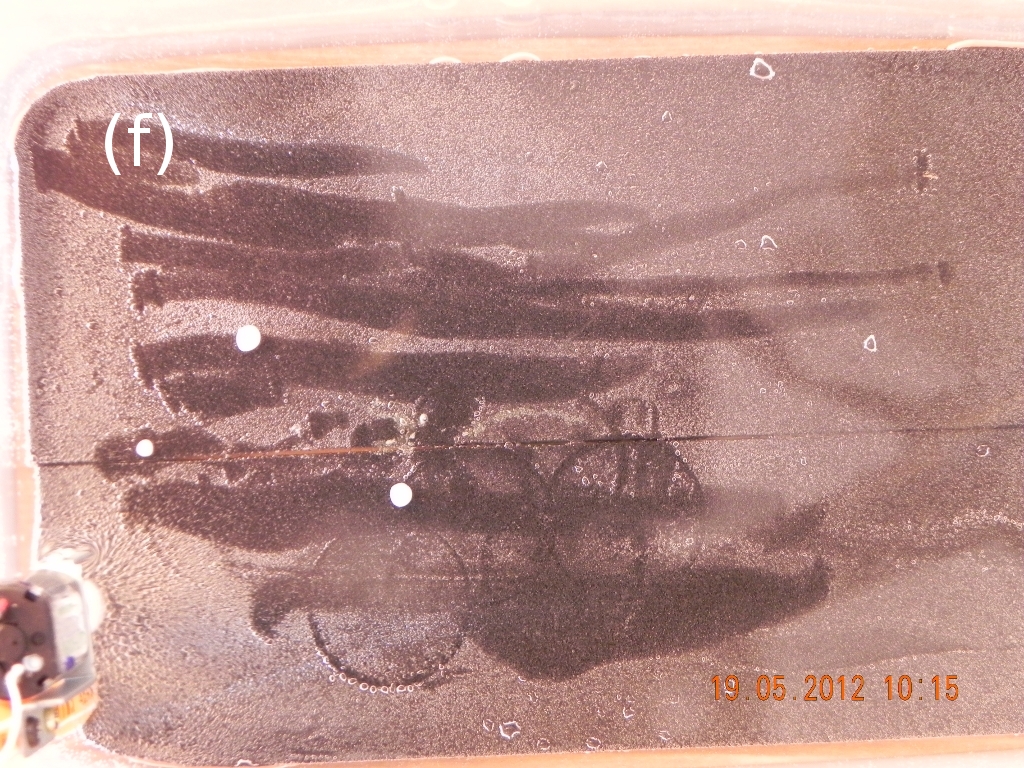}\\
\end{tabular}
\caption{A series of pictures taken throughout a certain measurement. (a) calibration with a 50 Euro cent coin; The dissolution of the candies after (b) 0, (c) 1, (d) 7, (e) 10 and (f) 15 minutes. The stripes are coming from the rubber background. The candies change their position slightly in time due to the flow provided by the stirring motor.}
\label{fig4}
\end{figure}

The calibration shot, as well as the on-going dissolution process of a certain measurement is shown in Figure \ref{fig4}.
From Figure \ref{fig4} it follows that under this experimental conditions the assumption that the candies maintain their spherical shape holds quite well. Of course, there were small deviations form the ideal sphere, but in average the shape was in good approximation indeed spherical. The diameter of the candies in terms of pixels was determined manually by using the freely available image manipulation program \texttt{GIMP} \cite{GIMP}. Furthermore, an estimate of the error of the diameter is provided by the number of pixels situated between the last one definitely belonging to the candy and the first one belonging definitely to the black bottom of the bowl, accounting for the uncertainty of where the actual rim of the candy is located. 
\section{\label{sec:results}Results}
\begin{figure}[h]
\centering
\includegraphics[width=10cm]{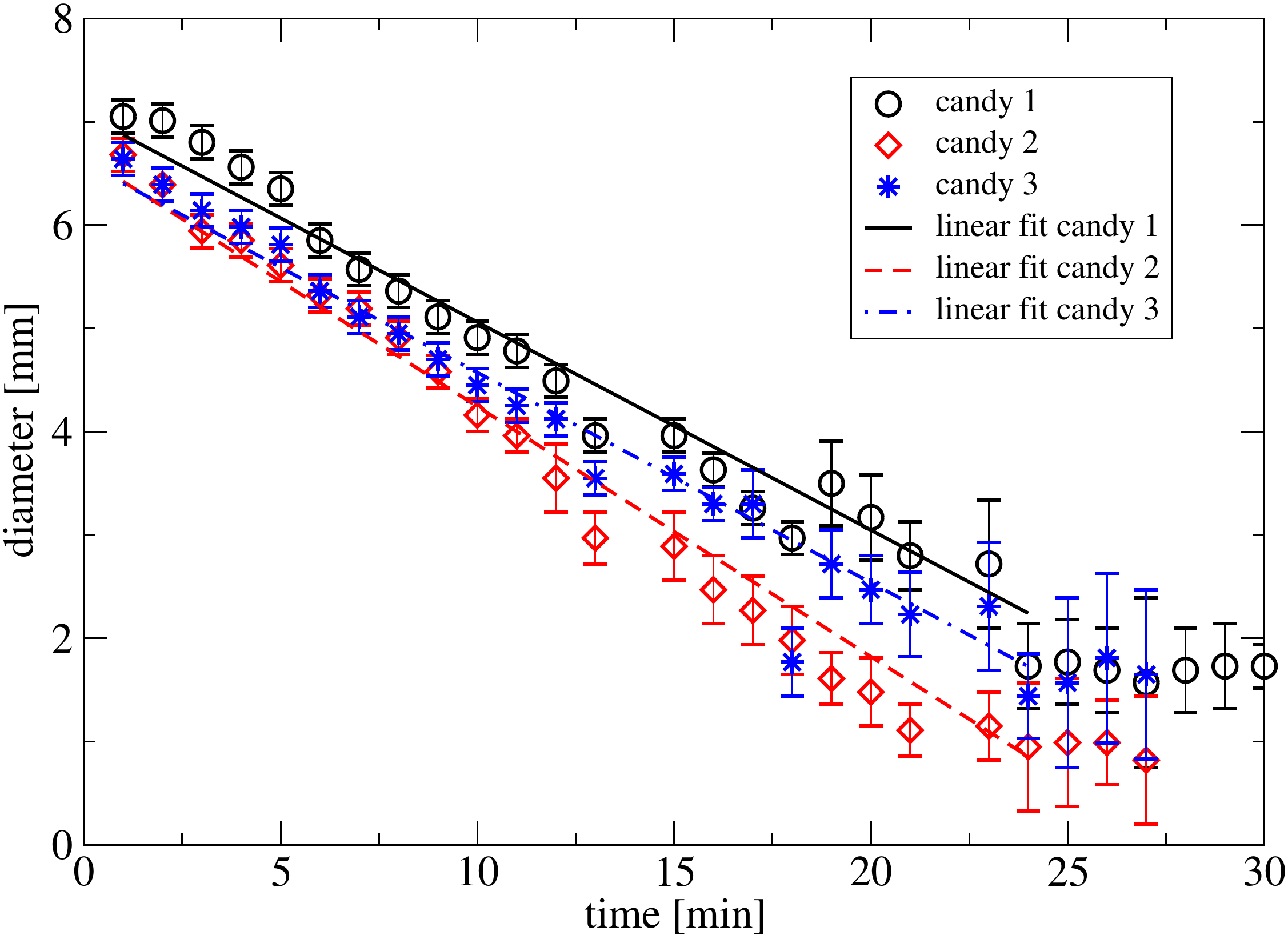}
\caption{The result of the measurement depicted in Figure \ref{fig4}. The three candies dissolve more or less linearly in time, as proposed by our simple model. Below $2\ mm$ in diameter, the behavior of the dissolution changes drastically. These points have been excluded from the fit-area.}
\label{fig5}
\end{figure}

Figure \ref{fig5} shows the results of the measurement depicted in Figure \ref{fig4}.
Above the $2\ mm$ diameter mark, the overall behavior of the candy-diameter as a function of time shows a linear decrease in good approximation. By applying a linear regression to these data points we get the following fit-parameters for the diameter $d$: $d = 7.0732 - 0.20123\ t$ with -0.986157, $d = 6.6637 - 0.24202\ t$ with -0.9867724 and $d = 6.6049 - 0.20325\ t$ with -0.980611 for the fit-function and correlation coefficient for candy 1, 2 and 3 respectively. With  correlation coefficients of more than 98\% we get a strong correlation for the linear decrease for all 3 candies. 
\section{Discussion}
In this study we raised, investigated and answered the question on how spherical candies dissolve in time. Providing a rather simple model for the dissolution process we claimed that such a candy should vanish linearly when characterized by its diameter. After the derivation and discussion of the model we constructed an experimental setup which allowed us to investigate the dissolution process of the candies quite close to the model's assumptions. We found good agreement with our model above roughly $2\ mm$ in diameter. Below the $2\ mm$ diameter mark we considered the existence of a core with a different density as a possible explanation for the deviation from the linear decrease of the diameter, even though the candy looks like it is homogeneous when broken apart. As finally confirmed by the manufacturer of the candy-samples, a core with a different density is used in the production process of the candies, which validated our assumption. The outer material of the shell is \textit{sugar-coated} onto this core. However, above the core-radius, our model produces a valid prediction in an easy-to-follow approach.\par 
\section{Conclusion}
Finally we would like to address the question proposed in the very beginning of this study: What is the best strategy of eating such a candy? As so often, the answer depends on what the person enjoying the candy considers as the optimum. If the time the candy lives should be maximized, the eater of the candy should try to maintain the spherical shape of the candy by all costs. Since the effect of mass transfer is driven by the surface, and the sphere possesses the smallest surface for a given volume among all possible shapes \cite{Steiner}, any deviation of the spherical shape increases the process of losing mass. In particular, breaking the candy with the teeth enlarges the surface by a huge amount, making the candy vanish faster. Thus, from this point of view one should carefully try to keep the candy as spherical as possible. But there is another way to look at it: Suppose you break the candy with your teeth in many pieces. The surface becomes big, and in an instant the mass that is transferred away from the fragments becomes huge as well. This might amplify the effect of tastiness and joy, even though the life-time of the candy has become considerably short in this approach. Even though we now \textit{know} how candies dissolve in time we stress that the best thing to do when eating a candy is to forget about these considerations, since they draw your attention away from what candies are made for: enjoyment.
\section*{Acknowledgments}
We thank Elvira Ortlepp of \textit{Schokoladenwerk Berggold GmbH} for the kind answers to our questions. Furthermore, we would like to thank Prof. Leopold Mathelitsch, Markus Hopfer and Prof. Reinhard Alkofer from the University of Graz for their careful reading of the manuscript. Financial support of the FWF Doctoral school, FWF DK W1203-N16, is acknowledged.

\end{document}